\documentstyle[12pt]{article} 
\renewcommand{\theequation}{\thesection-\arabic{equation}}
\bibliography{plain}
\title{\bf Effects of polydispersity on the phase coexistence diagrams in multiblock copolymers with Laser block  length distribution
} \vspace{20mm} \author{M.
Foroutan$^{a,c}$\thanks{E-mail:m.foroutan@tabrizu.ac.ir}$\;$ and
M. A.
Jafarizadeh$^{b,c,d}$\thanks{E-mail:jafarizadeh@tabrizu.ac.ir} .\\
$^a${\small Department of Physical Chemistry, Faculty of
Chemistry, Tabriz University, Tabriz 51664, Iran.}
\\ $^b${\small Department of Theoretical Physics and Astrophysics, Tabriz University, Tabriz 51664, Iran.}
\\ $^c${\small Institute for Studies in Theoretical Physics and Mathematics,
Teheran 19395-1795, Iran.}
\\ $^d${\small Research Institute for Fundamental Sciences, Tabriz 51664, Iran.}}
\newpage
\begin{document} \maketitle \vspace{15mm}
\newpage
\begin{abstract}
 Phase behavior  of $\mathbf{AB}$-multiblock copolymer melts which
consists of chains with  Laser distribution of $\mathbf{A}$ and
$\mathbf{B}$ blocks have been investigated in the framework of the
mean-field theory, where the polydispersity  of copolymer is a
function of two parameters $\mathbf{K}$ and $\mathbf{M}$. The
influence of the Laser distribution on higher order correlation
functions ( up to sixth order ) are computed for various values of
$\mathbf{K}$ and $\mathbf{M}$, and their contributions on  the
phase diagrams and phase coexistence  are presented. It is shown
that, with increasing polydispersity ( decreasing $\mathbf{K}$ and
increasing $\mathbf{M}$ ) the transition lines of all phases shift
upwards, consequently polydispersity destabilize the system.
 {\bf PACs
numbers:05.45.Ra, 05.45.Jn, 05.45.Tp }
\end{abstract}
\pagebreak \vspace{7cm}
\section{ Introduction}
Recently several publications have been appeared in connection
with block copolymer systems from both theoretical
\cite{Bind,Mats,Kiel} and experimental \cite{Rage,Zhan,Lowe}
aspects. In $\mathbf{AB}$ multiblock copolymer melts, the
immiscibility between $\mathbf{A}$ and $\mathbf{B}$ blocks induces
self-assembly into various ordered microstructures. With
decreasing the temperature, multiblock copolymers with repulsive
interaction between different monomers undergo a microphase
separation transition manifesting an energetic preference for
contacts between similar monomers.  In this work , the system will
be studied in the weak segregation regime, where the separation
between the $\mathbf{A}$ and $\mathbf{B}$ monomers is not
complete. The theoretical investigations demonstrate that it is
very possible that these (microphases separated) single phase are
metastable and the free energy can be lowered even more by
splitting the system into coexistence phase \cite{Nesa}.  \ The
properties of copolymer depend not only on the nature of the
comonomers and the overall compositions, but also on the
distribution of monomer units along the chain. As an example one
can consider N-vinyl pyrrolidone (VP) and vinyl acetate (VA)
copolymer. This copolymer can be synthesized in different ways
resulting in the different homogeneity of backbone monomer
distribution. Experimental studies show that the monomer unit
distribution has effect on solution properties,particularly when
the mole ratio VP to VA in the copolymer is close to unity,
sequence distribution plays an important role in solution behavior
\cite{Zhon}. In this work as in  Reference \cite{f},  we use the
most comprehensive distribution, namely Laser distribution with
two parameters $\mathbf{K}$ and $\mathbf{M}$ as the controller of
 degree of polydispersity. Polydispersity is an important
parameter which can effect many physical properties of a polymer.
Recently in an experimental work, the polydispersity lower than
1.0 is reported \cite{Adam}. Also the theoretical models with
polydispersity values of lower than unity can be useful  for
molecular weight distributions based on complex numbers
\cite{Eibe}.
\\The aim of the present paper is to investigate the
effect of the parameters of the Laser distribution on the position
of the phase transition lines, the stability of the various
microstructre and the position of the phase coexistence
 in multiblock copolymer.

  We consider an incompressible multiblock
copolymer melt in which both the lengths of $\mathbf{A}$-blocks
and $\mathbf{B}$-blocks satisfy Laser distribution. The phase
behavior of the multiblock copolymer melt is calculated in the
mean-field theory. By using this theory it is shown that there are
regions in the phase diagrams where two phases with different
symmetry of the superlattice (lamellar, hexagonal and bcc) coexist
with each other in the finite temperature range. Since the block
lengths in the system are assumed very large, this approximation
becomes more accurate and we can ignore the fluctuation
corrections \cite{Fred}.\\ The paper is organized as follows: In
section $\mathbf{II}$, the model and the block length distribution
are explained in details. Using the block length distribution one
can calculate the correlation functions which are required for
determination of the vertex functions of Landau free energy of
multiblock copolymer. In section $\mathbf{III}$ we calculate the
higher correlation functions up to sixth order and then  the
effects of the Laser distribution and Schultz distribution on them
are compared. In section $\mathbf{IV}$ we give a general
expression for the Landau free energy of a broad class of
polydisperse multiblock copolymers.  In Section $\mathbf{V}$ the
influence of the parameters $\mathbf{M}$ and $\mathbf{K}$ on phase
diagrams of $\mathbf{AB}$ multiblock copolymers are discussed. The
paper ends with conclusion and two appendices.

\section{ Laser distribution and its polydispersity }
We consider an incompressible multiblock copolymer melts
consisting of two types   of blocks $\mathbf{A}$ and $\mathbf{B}$
in which length of both $\mathbf{A}$  and B blocks obeys Laser
distribution. Generally the block length distribution is the
statistical representation of the probability of attachment of one
block to another block. The Laser distribution is used for the
first time as a comprehensive block length distribution in
multiblock copolymers  in Reference\cite{f}, such that it can be
reduced to other distributions in special cases. The normalized
Laser distribution is defined as \cite{Carr}: $$ P(n) =\frac{
K(M+1)^{\frac{K+1}{2}}}{n_n M}(\frac{
n}{n_n})^{\frac{K-1}{2}}\exp({ \frac{-K}{M}- \frac{
Kn(M+1)}{n_nM}}) I_{K-1}({
\frac{2Kn^{\frac{1}{2}}(M+1)^{\frac{1}{2}}}{n_n^{\frac{1}{2}}M}}),
$$ where $K >0$ and $I_{k-1}$ is modified Bessel function. For
$K=1 $, the Laser distribution reduces to the Laguerre
distribution. As $M\longrightarrow\infty $, the Laser distribution
reduces to Schultz-Zimm distribution\cite{Schu,Zimm}. Finally for
$K = 1$ and in the limit of $M \longrightarrow\infty$, the Laser
distribution reduces to the Flory distribution. As it will be
shown in Appendix $\mathbf{I}$, the polydispersity of the Laser
distribution is equal to $\frac{M(M+2)}{ K(M+1)^2} $. Clearly as
$\mathbf{M}$ becomes infinite, its polydispersity reduces to
$\frac{1}{K}$.
\section{ Higher correlation functions}
In this section, we calculate the higher correlation functions,
that is, the fourth, the fifth and the sixth order correlation
functions. The calculation of the lower correlation function are
given in details in Appendix II. The higher order correlation
functions are defined as: $$
g_{AAAA}=4!(n_A+n_B)^{3}(-\frac{3(1-\alpha)}{y^4} +\frac{f-2
{\alpha}^{\prime}}{y^3}+\frac{ {\alpha}^{\prime\prime}}{2y^2}) $$
$$g_{AAAAA}=5!(n_A+n_B)^{4}(\frac{4(\alpha-1)}{y^5}+\frac{f-3
{\alpha}^{\prime}}{y^4}
+\frac{{{\alpha}^{\prime\prime}}}{y^3}-\frac{{\alpha}^{\prime\prime\prime}}{6y^2})
$$
$$g_{AAAAAA}=6!(n_A+n_B)^{5}(\frac{3(\alpha-1)}{y^6}-\frac{f+4{\alpha}^{\prime}}{y^5}+\frac{3{\alpha}^{\prime\prime}}{2y^4}-\frac{
{\alpha}^{\prime\prime\prime}}{3y^3}+\frac{
{\alpha}^{\prime\prime\prime\prime}}{24y^2}) $$with
\begin{eqnarray}\nonumber
\alpha^{(i)}= (\frac{-1}{n_A+n_B})^i\int dn n^{i}\exp(-nx)
P(n)\end{eqnarray}
 where  $\alpha^{(i)}$ corresponds to
$\alpha$,$\alpha^{\prime}$,$\alpha^{\prime\prime}$,$\alpha^{\prime\prime\prime}$,$\alpha^{\prime\prime\prime\prime}$
for i=0,1,2,3,4, respectively . As it is  shown in the above
equation, the value of $\alpha^{(i)}$depends on the distribution
function P(n) where here we use the most comprehensive
distribution, that is, Laser distribution, where it can be reduced
to Schultz distribution for paticular values of its parameters. In
order to obtain the correlation functions, we have to calculate
the Laplace transforms of $\alpha(y)$. The Laplace transforms of
$\alpha(y)$ (corresponding to Laser distribution) are: $$
\alpha^{(i)}=R! i! \beta^{k-1}
 \gamma^{-(k+i)}L_{i}^{(k-1)}(-\frac{\beta^2}{\gamma})$$
$$R=\frac{k(M+1)^{\frac{k+1}{2}}}{Mn_n}
(\frac{1}{n_n})^{\frac{k-1}{2}}\exp{(\frac{-k}{M})}$$ $$
\gamma=\frac{k(M+1)+Mn_{n}}{Mn_{n}} $$ and
$$\beta=\frac{k(M+1)}{Mn_{n}^\frac{1}{2}},$$ where the expression
$L_{i}^{(k-1)}$ is the Laguerre function. The influence of the
parameters $\mathbf{K}$ and $\mathbf{M}$ on the shape of the
second, the third and the sixth order correlation functions are
given in the figures 1a to 1c. The form of the third, fourth,
fifth and sixth order correlations are same but their scales are
very different( they are of order: $10^{9}$, $10^{15}$, $10^{21}$
and $10^{27}$, respectively). If we apply the Schultz
distribution, we obtain only the first point of the right hand of
the curves. It should be reminded that as $M\longrightarrow\infty
$, the Laser distribution reduces to Schultz-Zimm distribution.
\section{ Theory of Landau free energy }
 From the point of view of Landau's phase transition theory \cite{Land}, a
system under microphase  separation undergoes a set of phase
transitions  which are described in terms of the order parameter:
$$ \psi(x)=\rho_A(x)-<\rho_A>, $$
 where $\rho_A(x)$ is the local density of $\mathbf{A}$ monomers and  $ < \rho_A>$ is the average
value of the density over the volume $\mathbf{V}$ of the system.
At low temperature, the order parameter can become a spatially
periodic function, which possesses  the symmetries of a certain
space group $\zeta$ i.e. : $$ \psi(\vec{q}) =\Sigma_{\vec{Q}\in
\hat{L}}A_{\vec{Q}} \exp{( i\Phi_{\vec{Q}})}\delta
(\vec{q}-\vec{Q}),$$ where $\psi(\vec{q})$ is the Fourier
transform of concentration profile $\psi(x)$, $\hat{L}$ is the
reciprocal of the lattice L describing the translational
symmetries of $\zeta$, the amplitudes $A_Q$ and phases $\Phi_Q$
are adjustable parameters. The  free  energy of the system is : $$
F ( \chi , f )  =  - K_B T  \ln Z ( \chi , f ), $$ where  $T$ is
temperature, $ K_B$  is  Boltzmann's   constant, $\chi$ is the
Flory-Huggins parameter, $\mathbf{f} $ is  volume fraction of the
$ A$ block and  $Z $  is  the  partition function defined as the
integral over all possible profiles $\psi(x)$, i.e. : $$ Z = \int
d\psi \exp{(-H(\psi(x)) / K_B T }),$$ where the effective
Hamiltonian, $ H(\psi(x))$ is the virtual free energy of the state
with a given $\psi(x)$. Since we assume that $\rho_A(x)$ differs
from $<\rho_A>$ only slightly, the free energy for each symmetries
of the superlattice i may be approximated by its Landau's
expansion  in powers of $\psi(x)$, i.e. : $$ F_L^i (\psi(x))= K_B
T\sum \frac{1}{n!}\int dx_1\cdots dx_n \Gamma _{n}( x_1\cdots x_n
) \psi(x_1)\cdots \psi(x_n), $$ with vertices $\Gamma_n$ defined
as :
\begin{eqnarray}\nonumber
\Gamma_2({{\overrightarrow{q_1}},{\overrightarrow{q_2}}})=V\delta_k
({\overrightarrow{q_1}},{\overrightarrow{q_2}})[\frac{g_{AA}(q_1)+2g_{AB}(q_1)+g_{BB}(q_1)}{g_{AA}(q_1)*g_{BB}(q_1)-g_{AB}^2(q_1)}-2\chi],
 \nonumber
\end{eqnarray}
\begin{eqnarray}\nonumber
\Gamma_3({{\overrightarrow{q_1}},{\overrightarrow{q_2}},{\overrightarrow{q_3}}})=-V\delta_k
({\overrightarrow{q_1}},{\overrightarrow{q_2}},{\overrightarrow{q_3}})\sum
g_{\alpha\beta\gamma
}({\overrightarrow{q_1}},{\overrightarrow{q_2}},{\overrightarrow{q_3}})z_{\alpha}(q_1)
z_{\beta}(q_2)z_{\gamma}(q_3), \nonumber
\end{eqnarray}
where $g_{\alpha\beta}$ and $g_{\alpha\beta\gamma}$ are second and
third correlation functions, respectively, and
\begin{eqnarray}\nonumber
z_{\alpha}(q)=g_{\alpha A}^{ -1}(q)- g_{\alpha
B}^{-1}(q).\nonumber \end{eqnarray} After inserting  the trial
function $\psi(x)$ in Landau free energy, the lowest order term
$F_2$ attains the form: $$ \frac{F_2}{V}=\gamma _2 A^{2}, $$ This
expression is independent of the structure. The third order and
fourth order contribution to free energy are different for the
various structures:
\begin{eqnarray}
\nonumber \frac{F_3 ^{hex}}{V}   = -\frac{2 \mid\gamma _3\mid
A^{3}}{ 3 \sqrt{3}}, \nonumber\\  \frac{F_3 ^{bcc}}{V}  = \frac{-4
\mid \gamma _3\mid A^{3 }}{ 3 \sqrt{6}}, \nonumber \\
 \frac{F_3 ^{lam }}{V} = 0, \nonumber\end{eqnarray}\\
while the fourth  order contribution to the free energy has the
form
$$\frac{1}{4!}\sum_{{\overrightarrow{q_1}},{\overrightarrow{q_2}},
{\overrightarrow{q_3}},{\overrightarrow{q_4}}}
\Gamma_4({\overrightarrow{q_1}},{\overrightarrow{q_2}},{\overrightarrow{q_3}},{\overrightarrow{q_4}})
A^4.$$
 In order to obtain the vertices of multiblock
copolymer whose blocks lengths obey a certain distribution
$\mathbf {P(n)}$, we have to calculate the correlation functions.
The general expressions for the correlation functions are
investigated in Appendix II. The phase diagram can be obtained by
numerical minimization of free energy with respect to the
parameters $\mathbf{A}$ and $\mathbf{y}$. Several phases with
different superlattice structure can coexist with each other.
Considering a two-phase state with the volume fractions
$\mathbf{x}$ and $\mathbf{(1-x)}$ occupied by phases 1 and 2,
respectively, the final expression for the free energy can be
defined as : $$ F=x
F_{1}(A_{1},y_{1})+(1-x)F_{2}(A_{2},y_{2})-\frac{1}{2}x(1-x)J $$\
and
$$J=4(\frac{(1}{y_{1}}+1)A_{1}^{4}+8(\frac{(1}{y_{1}+y_{2}}+1)A_{1}^{2}A_{2}^{2}-4(\frac{(1}{y_{2}}+1)A_{2}^{4}.$$
Here the phase diagram can be obtained by numerical minimization
of free energy with respect to the parameters $\mathbf{A_1}$,
$\mathbf{A_2}$, $\mathbf{y_1}$ and $\mathbf{y_2}$ and $\mathbf{x}$
for each coexistence.

\section{The influence of the polydispersity on the phase diagrams and phase coexistence }
In this section phase coexistence for various values of
polydispersity with Laser distribution are presented in the
framework of the mean-field theory. Here  diagrams  in the
immediate vicinity  of the critical point of the multiblock
copolymers for $\mathbf{K}=1$ to 40 and $\mathbf{M}=0.01$ to 100
are constructed. Figure 2 shows the mean-field phase diagrams for
the most polydispers and the most monodisperse systems,
respectively. In figure $2a$, $\mathbf{M}$ is very large and
$\mathbf{K}$ is very small and in figure $2b$, $\mathbf{M}$ is
very small and $\mathbf{K}$ is very large. Figures 3 and 4 show
the mean-field phase diagram for the multiblock copolymer melts in
terms of variables $\mathbf{N}{\chi}$, where
 ${\chi}$ is the interaction parameter and $\mathbf{f}$ is the monomer fraction. In
these figures system is at very  high polydisperse and very low
polydiperse states , respectively . It should be reminded that
according to the equation of the polydispersity of Laser
distribution, the polydispersity increases with increasing of
$\mathbf{M}$ and decreasing of $\mathbf{K}$. These figures show
that the region of the stability of the phase coexistence shift
considerably to the larger $\mathbf{N}{\chi}$  with increasing
polydispersity. Figures 5 and 6 show the one-dimensional phase
diagrams of multiblock copolymer with $\mathbf{f}=0.25$ for
$\mathbf{K}=1$ and $\mathbf{K}=40$, respectively. One can see that
for low values of $\mathbf{K}$, the value  of $\mathbf{M}$ has not
considerable effect on the stability region of microphases while
its effect  is very large for higher values of $\mathbf{K}$.
Therefore, for large values of $\mathbf{K}$, with the decreasing
of $\mathbf{M}$, the transition lines of the phases shifts
severely upwards, while for small values of $\mathbf{K}$ these
changes are very slow. Great attention should be paid to the
scales in this figure. Even though these scales are unphysical,
but it can be very useful in showing the sensitivity of
$\mathbf{N}{\chi}$  to ${M}$. The dependence of the
$\mathbf{N}{\chi}$ on polydispersity, in the high polydisperse and
low polydisperse systems, with $\mathbf{f}=0.25$, are given in
figures 7 and 8, respectively. These figures show that for small
values of $\mathbf{K}$ with increasing of $\mathbf{M}$ the slope
of the curve is slow, while for large values of $\mathbf{K}$, it
is very sharp.\\ CONCLUSION :\\ In the present paper we have
introduced the most comprehensive distribution function for the
polydispersity of molecular weight distributions that is the Laser
distribution. Laser distribution with two parameters $\mathbf{K}$
and $\mathbf{M}$ can be reduced to other distributions. The Flory
distribution and Schultz-Zimm distribution  are
$\mathbf{M}\longrightarrow\infty$ limiting case of Laser
distribution, therefore, so far the role of finite values of
$\mathbf{M}$ has not been considered. Actually in this work we see
that the effect of $\mathbf{M}$ (especially for the low
$\mathbf{M}$) on the curve of distribution versus block length and
also its effect on the phase behaviour of multiblock copolymer is
very interesting.\\ \vspace{10mm} {\large
\appendix{Appendix I : The polydispersity of distributions}}\\
\renewcommand{\theequation}{\Roman{section}{I}-\arabic{equation}}
\setcounter{equation}{0}Here in this appendix we derive the
polydispersity of distribution functions. According to section III
the polydispersity is defined as : $$U =  \frac{< n^2> -
<n>^2}{<n>^2},$$ where $$<n^i> = \int dnP(n)n^i.$$ Using the above
formulas we calculate the polydispersity of Laser distribution
which is comprehensive  enough to include the other known
distributions as a limiting cases. Hence we can obtain the
polydispersity of other distributions from this one. In
calculating polydispersity of Laser distribution we have to use
the following integral: $$\int_{0}^{\infty} dx
x^{p+\upsilon/2}\exp(-\alpha x) I_\upsilon{(2\beta
x^{\frac{1}{2}})}=p! \beta ^\upsilon \exp{(\beta^2/\alpha)}\alpha
^{( -p-\upsilon -1)} L_\upsilon^p{(-\beta^2/\alpha)}$$. Using the
above integral we obtain the following expression for the p-moment
of laser distribution: $$<x^p> = p!\alpha^{-p}L_p^{k-1}(-\beta
^2/\alpha),$$where
$$L_p^\upsilon(x)=(1/p!)x^{-\upsilon}\exp(x)(d/dx)^p(x^{\upsilon+p
}\exp{(-x)} )  $$ is the Laguerre polynomial of order p. Now,
using the first and second moments we obtain the following
expression for polydispersity of Laser distribution: $$U =\frac{
M(M+2)}{K(M+1)^2}.$$ Obviously for $K = 1$ we get the Laguerre
distribution with polydispersity $U=1-\frac{1}{(M+1)^2}$ and as
${\mathbf M}\rightarrow\infty$ we obtain the  Schultz-Zimm
distribution with polydispersity $ U=\frac{1}{K}$ where   Flory
distribution with polydispersity $U = 1$ can be obtained from it
for $K=1$.
\\\\{\large
\appendix{Appendix II: The Correlation Functions of Multiblock
Copolymer \cite{Ange}}}
\renewcommand{\theequation}{\roman{section}{I}-\arabic{equation}}
\setcounter{equation}{0}\\ In a polydisperse multiblock copolymer
melt several molecule types are presented. if $\mathbf{s}$ denotes
a molecule type (i.e. a finite sequence of $\mathbf{A}$-blocks and
$\mathbf{B}$-blocks) and if $\mathbf{\rho_{s}}$ denotes the number
density of molecules of type s, then the composition of the melt
is fixed completely by the set $\mathbf{{\rho_{s}}}$. A molecule s
is fixed if the lengths of all blocks in the molecule are fixed:
$$ s=(N_A^{1},N_B^{1},N_A^{2},N_B^{2},\ldots,N_A^{Q},N_B^{Q})$$
where $N_\alpha^{i}$ denotes the length of the $i^{th}$block of
type $\alpha$ in the molecule. In  this work we have assumed that
the number of molecule (Q) is very large. Here $P_\alpha(n)$
denotes the probability that a block of type $\alpha$ has a length
between n and n+dn. For this distribution we have :
$$\int_{0}^{\infty}dnP_\alpha(n)
=1$$and$$\int_{0}^{\infty}dnP_\alpha(n)=n_\alpha,$$ where
$n_{\alpha}$ is the average length of a block of type $\alpha$.
The quantity $\rho_{s}$ can be defined as: $$
\rho_{s}=cP_{s}=cP_{A}(N_{A}^{1}) P_{B}(N_B^{1})P_{A}(N_{A}^{2})
P_{B}(N_B^{2})\ldots P_{A}(N_{A}^{Q}) P_{B}(N_B^{Q}).$$ The
normalization constant $\mathbf{c}$ is determined by the
constraint that the monomer density is unity, i.e. :
$$\sum_{s}\rho_{s}N_{s}=1,$$ where $N_{s}$ is the total molecule
length. Combining the above equations gives the normalization
constant c: $$c=\frac{1}{Q(n_{A}+n_{B})}.$$ Now we calculate the
second order function. The average second order function over the
composition ${\rho_{s}}$ is given by:
 $$g_{ \alpha\beta} =  \sum_{s}\rho _{s}\sum _{i,j}\sigma _{i}(\alpha)
\sigma _{j}(\beta)\exp(-q^2\mid j-i\mid)$$ where $\sigma_{
i}(\alpha)$ equals 1 only  if block i is of  type $\alpha$ and
otherwise is zero. Where i and j are many blocks apart, the weight
factor $ \exp(-q^2\mid{j-i}\mid)$ will give a negligible
contribution to the correlation function, and one gets a
contribution only if the number of blocks  between i and j is of
order unity. Here for simplicity, the calculation of the
correlation function where i and j are in the same block  are
given. For second correlation function, the
$\mathbf{AA}$-component of the second order correlation function
is equal to :\\
\begin{eqnarray}\nonumber g_{AA}=\frac{1}{Q(n_{A}+n_{B})}2Q\int dn P(n)\int_{0}^{n} di
\int_{0}^{n} dj \exp(-x(j-i)) =\nonumber\\ \frac{
2}{n_{A}+n_{B}}\int dn P(n){{\frac{n}{x} + \frac{\exp(-n x)}{x^2}
- \frac{1}{x^2}}} = (n_{A}+n_{B}) ( \frac{-2(1- \alpha (y))}{y^2}
+ \frac{2f}{y} )\nonumber\end{eqnarray}
 where:
\begin{eqnarray}\nonumber
 f = \frac{n_A}{n_A+n_B}\quad,x
=q^2\quad,y=(n_A+n_B)x ,
 \alpha(y)=\int dn\exp(-n x)P(n)\end{eqnarray}
 For the $\mathbf{BB}$-component of the second order correlation
function, $ \beta(y)$ is used to as the Laplace transform of  the
length distribution of the $\mathbf{B}$-blocks. Since the states
corresponding to $i<j$ give the same contribution as $j<i$ ones,
one should take into  account a factor 2 due to  this symmetry.
 Finally the factor Q can be interpreted as the number of blocks in
which monomer i can be presented.\ The contribution third order
correlation function for the situation where all three monomer are
in same block $\mathbf{A}$ is :
\begin{eqnarray}\nonumber g_{AAA}=\frac{1}{Q(n_{A}+n_{B})}3! Q
\int_{0}^{\infty} dn P(n)\int_{0}^{n} di \int_{i}^{n} dj
\int_{j}^{n} dk \exp(-x(k-i))\\\nonumber
=6(n_{A}+n_{B})^2(-\frac{2(1-\alpha)}{y^3}+\frac{f-\acute{\alpha}}{y^2})\end{eqnarray}
$\acute{\alpha}$ is the derivative of $\alpha$ with respect to y:
\begin{eqnarray}\nonumber
\acute{\alpha}(y)=d/dy\int_{0}^{\infty}dn\exp(-nx) P(n)\\
\nonumber  =-\frac{1}{(n_A+n_B)}\int_{0}^{\infty}dn n\exp(-nx)
P(n)
\end{eqnarray}

\newpage
Figures captions\\ FIG. 1.  Dependence of the second, third and
sixth order correlation functions on the value of $\mathbf{K}$ for
various values of $\mathbf{M}$, respectively (1a-1c).
\\ FIG. 2.Mean-field phase diagrams in the vicinity of the critical point
for multiblock copolymers for the most polydispers and the most
monodisperse system, respectively. The lower line separates the
disordered phase from the bcc phase, the middle line separates the
bcc phase from the hexagonal phase and the upper line separates
the hexagonal phase from the lamellar phase. $(2a)$ $\mathbf{K}=1$
and $\mathbf{M}=100$. $(2b)$ $\mathbf{K}=40$ and
$\mathbf{M}=0.01$.
\\FIG. 3. Mean-field phase diagram in the vicinity of the critical
point for multiblock copolymers with $\mathbf{K}=1$ and
$\mathbf{M}=100$. White areas denote single phase regions lam, hex
and bcc from up to down, respectively. Black areas denote the
regions of coexistence of 2-phases.
\\FIG. 4. Mean-field phase diagram  in the vicinity
of the critical point for multiblock copolymers with
$\mathbf{K}=40$ and $\mathbf{M}=0.01$. White areas denote single
phase regions lam, hex and bcc from up to down, respectively.
Black areas denote the regions of coexistence of   2-phases.
\\FIG. 5. The one-dimensional phase diagrams of the multiblock
copolymer melts $\mathbf{f}=0.25$ for the Laser distribution with
$\mathbf{K}=1$ for various values of $\mathbf{M}$.\\FIG. 6. The
one-dimensional phase diagrams of the multiblock copolymer melts
$\mathbf{f}=0.25$ for the Laser distribution with $\mathbf{K}=40$
for various values of $\mathbf{M}$.\\ FIG. 7. Dependence of the
parameter $\mathbf{N}{\chi}$ on the polydispersity from $M=0.01$
to $M\longrightarrow\infty$ for $K=1$. \\ FIG. 8.  Dependence of
the parameter ${N}{\chi}$ on the polydispersity from
$\mathbf{M}=0.01$ to $\mathbf{M}\longrightarrow\infty$ for
$\mathbf{K}=40$.
\end{document}